\documentclass{article}
\usepackage[pdfmenubar=false, letterpaper=false, pdfpagelabels=true]{hyperref}

\begin{document}

\title{The Fermi's Bayes Theorem}
\author{G. D'Agostini \\
Universit\`a ``La Sapienza'' and INFN, Roma, Italia \\
{\small giulio.dagostini@roma1.infn.it, \url{www.roma1.infn.it/~dagos}} \\
\mbox{} \\
{\it To appear in the Bulletin of the 
International Society of Bayesian Analysis}
}

\date{}
\maketitle

\begin{abstract}
It is curious to learn that Enrico Fermi knew how to base probabilistic
inference on Bayes theorem, and that some influential notes on statistics
for physicists stem from what the author calls elsewhere, 
but never in these notes, {\it the Bayes Theorem of Fermi}. 
The fact is curious because the large majority
of living physicists, educated in the second half of
last century -- a kind of middle age in the statistical reasoning --
never heard of Bayes theorem during their studies, though 
they have been constantly using an intuitive reasoning quite 
Bayesian in spirit.
This paper is based on recollections and notes by Jay Orear and 
on Gauss' ``Theoria motus corporum coelestium'',  being 
the {\it Princeps mathematicorum} remembered by Orear as source of 
Fermi's Bayesian reasoning.

\end{abstract}

\section*{}
Enrico Fermi is usually associated by the general public
with the first self-staining nuclear chain reaction
and, somehow, with the Manhattan Project to build the first atomic bomb. 
But besides these achievements, that set a mark in history,
his contribution to physics - and especially 
fundamental physics - was immense, as testified for example by 
the frequency his name,
or a derived noun or adjective, appears in the scientific literature
(fermi, fermium, fermion, F. interaction, F. constant, Thomas-F. model,
F. gas, F. energy, F. coordinates, F. acceleration mechanism, etc.).
Indeed he was one of the founding fathers of atomic, nuclear, 
particle and solid state physics, with some relevant contributions 
even in general relativity and astrophysics.

He certainly mastered probability theory and one of his chief
interests through his life was the study of 
the statistical behavior of physical systems of free or 
interacting particles. 
Indeed, there is a `statistics' that carries his name, together
with that of the co-inventor Paul Dirac, and the particles 
described by the Fermi-Dirac statistics are called fermions.

Among the several other contributions of Enrico Fermi to statistical
mechanics, perhaps the most important is contained in his last
paper, written with John Pasta and Stan Ulam.
Without entering into the physics contents of the paper
(it deals with what is presently known as the `FPU problem') 
it is worth mentioning the innovative
technical-methodological issue of the  work:
the time evolution of a statistical system (just a chain of 
nonlinearly coupled masses and springs)
was simulated by computer. The highly unexpected result
stressed the importance of using numerical simulations
as a research tool complementary to theoretical studies 
or laboratory experiments. Therefore, Fermi, who was unique 
in mastering at his level both theory and experiments,
was also one of the first physicists doing `computer experiments'. 

In fact, with the advent of the first electronic computers, Fermi
immediately realized the importance of using them to solve
complex problems that lead to difficult or intractable 
systems of integral-differential equations. One use of the
computer consisted in discretizing  the problem and 
solving it by numerical steps (as in the FPU problem). 
The other use consisted in applying sampling techniques, 
of which Fermi is also recognized to be a pioneer. 
It seems in fact, as also acknowledged by Nick 
Metropolis,\footnote{\url{http://library.lanl.gov/cgi-bin/getfile?00326866.pdf}}
that Fermi contrived and used the Monte Carlo method
to solve practical neutron diffusion problems in 
the early nineteen thirties, i.e. fifteen years before
the method was finally `invented' by Ulam, named by 
Metropolis, and implemented on the first electronic computer
thanks to the interest and drive of Ulam and John von Neumann.

After this short presentation of the character, with emphasis
on something that might concern the reader of this bulletin,
one might be interested about Fermi and `statistics', 
meant as a data analysis tool. During my studies and later
I had never found Fermi's name in the books and lecture notes on statistics
I was familiar with. It has then been a surprise to read the
following recollection of his former student Jay Orear,
presented during a meeting to celebrate the 2001 centenary 
of Fermi's birth: 
{\em ``In my thesis I had to find the best 3-parameter 
fit to my data and the errors of those parameters in order to get
the 3 phase shifts and their errors. Fermi showed me a simple analytic
method. At the same time other physicists were using and
publishing other cumbersome methods. Also
Fermi taught me a general method, which he called Bayes Theorem,
where one could easily derive the best-fit parameters and their
errors as a special case of the maximum-likelihood method''}

Presently this recollection is included in the freely 
available Orear's book
 ``Enrico Fermi, the master scientist''.\footnote{\url{http://hdl.handle.net/1813/74}}
So we can now learn that Fermi was teaching his students a maximum likelihood
method {\em ``derived from his Bayes Theorem''} and that 
{\em ``the Bayes Theorem of Fermi''} - so Orear calls it  - 
is a special case of Bayes Theorem, in which the priors are equally likely
(and this assumption is explicitly stated!). Essentially, 
Fermi was teaching his young collaborators to use likelihood ratio
to quantify how the data preferred one hypothesis among several possibilities,
or to use the normalized likelihood to perform 
parametric inference (including the assumption of Gaussian
approximation of the final pdf, that simplifies the calculations).

Fermi was, among other things, an extraordinary teacher,
a gift witnessed by his absolute record in number of 
pupils winning the Nobel prize  - up to about a dozen, 
depending on how one counts them. But in the case of probability
based data analysis, it seems his pupils didn't get fully 
the spirit of the reasoning and, when they remained orphans 
of their untimely dead scientific father, they were in an uneasy
position between the words of the teacher and the dominating 
statistical culture of those times. Bayes theorem, and especially
his application to data analysis, appears in  Orear's book 
as one of the Fermi's working rules, of the kind of the 
`Fermi golden rule' to calculate reaction probabilities. 
Therefore Orear reports of his ingenuous question to know 
{\em ``how and when he learned this''} (how to derive maximum
likelihood method from a more general tool).
Orear {\em ``expected him to answer R.A. Fisher or some textbook 
on mathematical statistics''}. 
{\em ``Instead he said, `perhaps it was Gauss'\,''}.
And, according to his pupil, Fermi {\em ``was embarrassed to 
admit that he had  derived it all from his Bayes Theorem''}. 

This last quote from Orear's book gives an idea of the 
author's unease with that mysterious theorem and of his reverence 
for his teacher: {\em ``It is my opinion that Fermi's statement of Bayesian
Theorem is not the same as that of the professional mathematicians
 but that Fermi's version is nonetheless simple and powerful. Just
as Fermi would invent much of physics independent of others, so 
would he invent mathematics''}. 

Unfortunately, Fermi wrote nothing on the subject. 
The other indirect source of information we have are the 
 ``Notes on statistics for physicists'', written by Orear
in 1958, where the author acknowledges that his {\em ``first introduction
to much of the material here was in a series of discussions
with Enrico Fermi''} and others {\em ``in the autumn 1953''}
(Fermi died the following year). A revised copy of the notes
is available on the 
web.\footnote{See e.g. \url{http://nedwww.ipac.caltech.edu/level5/Sept01/Orear/frames.html}}

When I read the titles of the first two sections, ``Direct probability''
and ``Inverse probability'', 
 I was hoping to find there a detailed account of the Fermi's Bayes
Theorem. But I was immediately disappointed. Section 1 starts
saying that {\em ``books have been written on the `definition'
of probability''} and the author abstains from providing one, 
jumping to two properties of probability: statistical 
independence (not really explained) and the law of large numbers,
put in a way that could be read as Bernoulli theorem as well 
as the frequentist definition of probability. 

In Section 2, ``Inverse probability'', there is no mention to 
Bayes theorem, or to the Fermi's Bayes Theorem. Here we clearly see
the experienced physicist tottering between the physics 
intuition, quite `Bayesian', and the academic education
on statistics, strictly frequentist (I have written years
ago about this conflict and its harmful 
consequences.\footnote{\url{http://xxx.lanl.gov/abs/physics/9811046}} 
Therefore Orear explains {\em ``what the physicist usually means''}
by a result reported in the form `best value $\pm$ error': the physicist
{\em ``means the `probability' of finding''} 
{\em ``the true physical value of the parameter under question''}
in the interval `[best value - error, best value + error]' is
such and such percent. 
But then, the author immediately adds that  {\em ``the use of the word 
`probability' in the previous sentence would shock the mathematician''},
because {\em ``he would say that the probability''} the 
quantity is in that interval {\em ``is either 0 or 1''}. 
The section ends with a final acknowledgments of the conceptual difficulty
and a statement of pragmatism: {\em ``the kind of probability the physicist is
talking about here we shall call inverse probability, in contrast to the
direct probability used by the mathematicians. Most physicists use the same 
word, probability, for the two different concepts: direct probability
and inverse probability. In the remainder of this report we will
conform to the sloppy physics-usage of the word `probability' ''}.

Then, in the following sections he essentially  
presents a kind of hidden  Bayesian 
approach to model comparison (only simple models) and parametric
inference under the hypothesis of uniform prior, under which his
guiding Fermi's Bayes Theorem held.

Historians and sociologists of science might be interested in 
understanding the impact Orear's notes have had 
in books for physicists written in the last forty-fifty years,
and wonder how they would have been if the word 'Bayes' 
had been explicitly written in the notes. 

Another question, which might be common to many readers at this point,
is why Fermi associated Gauss' name to Bayes theorem. I am not familiar
with all the original work of Gauss and a professional historian would 
be more appropriate. Anyway, I try to help with the little I know. 
In the derivation of the normal distribution (pp. 205-212 of his 1809 
``Theoria motus corporum coelestium in sectionibus
conicis solem ambientum'' -- I gave a short account of these
pages in a book), 
Gauss develops a reasoning to invert the
probability which is exactly Bayes theorem for hypotheses that 
are {\em a priori} equally 
likely\footnote{Something similar, 
also independently from Bayes, 
was done by Laplace in 1774 (see Stephen Stigler's
`The History of Statistics'). However Gauss does not mention
Laplace for this result in his 1809 book (while, instead,
he acknowledges him for the integral to normalize the Gaussian!). 
Therefore the `Fermi's Bayes Theorem' should be, more properly,
 a kind of `Laplace-Gauss Theorem'.}
(the concepts of prior and posterior are well stated by Gauss), 
and, later, he extends
the reasoning to the case of continuous variables. That is 
essentially what Fermi taught his collaborators. 
But Gauss never mentions Bayes, 
at least in the cited pages, and the use of the `Bayesian'
reasoning is different from what we usually do: 
we start from likelihood and  prior (often uniform or quite `vaque')
to get the posterior. Instead, Gauss got a general form 
of likelihood (his famous error distribution) 
from some assumptions: uniform prior; same error
function for all measurements;  some analytic
property of the searched-for function;  
posterior maximized at the arithmetic average of data points.

Then, why did Fermi mention Gauss for the name of the theorem 
and for the derivation of the maximum likelihood method from the 
theorem? Perhaps he had in mind another work of 
Gauss. Or it could be --
I tend to believe more this second hypothesis -- 
a typical Fermi unreliability in providing references,
like in the following episode 
reported by Lincoln Wolfenstein in his contribution to Orear's book:
{\em ``I remember the quantum mechanics course, where students would
always ask, `Well, could you tell us where we could find that in 
a book?' And Fermi said, grinning, `It's in any quantum mechanics
book!' He didn't know any. They would say, `well, name one!' 
`Rojanski', he said, `it's in Rojanski'. 
Well, it wasn't in Rojanski -- it wasn't in any quantum mechanics
book.''}  

I guess that, also in this case, most likely {\it it wasn't in Gauss}, 
though some seeds {\it were} in Gauss. In the pages that immediately
follow his derivation of the normal distribution, Gauss shows
that, using {\it his} error function, with the same 
function for all measurements, the posterior is maximized when 
the sum of the squares of residual is minimized. 
He recovered then the already known
least square principle, that he claims to be his principle 
({\it ``principium nostrum''}, in Latin) used since
1795, although he 
acknowledges Legendre to have published a similar principle in 1806. 
Therefore, since Gauss used a flat prior, his `Bayesian' derivation of the
least square method is just a particular case of the maximum likelihood
method. Fermi must have had this in mind, together with
Bayes' name from modern literature and with many logical consequences 
that were not really in Gauss, when he replied young Orear. 

\mbox{ }

[\,Some interesting links concerning this subject, including pages 
205-224 of Gauss' ``Theoria motus corporum coelestium'', can be found
in \url{http://www.roma1.infn.it/~dagos/history/}.\,]

\end{document}